\documentclass{elsart}
\usepackage{graphicx}
\usepackage{amssymb}
\usepackage{amsmath}
\begin{document}
\begin{frontmatter}
\title{Electronic structure, electron-phonon coupling and 
superconductivity of isotypic noncentrosymmetric crystals 
Li$_2$Pd$_3$B and Li$_2$Pt$_3$B}
\author{S.\ K.\ Bose}
\ead{bose@brocku.ca} and
\author{E.\ S.\ Zijlstra}
\address{Physics Department, Brock University, St.\ Catharines, 
Ontario, Canada L2S 3A1}

\begin{abstract}
Electronic structure of recently discovered isotypic ternary borides 
Li$_2$Pd$_3$B and Li$_2$Pt$_3$B, with noncentrosymmetric crystal
structures, is studied with a view to understanding their 
superconducting properties. 
Estimates of the Fermi-surface averaged electron-phonon matrix element
and Hopfield parameter are obtained in the rigid ion approximation of 
Gaspari and Gyorffy [Phys.\ Rev.\ Lett.\ {\bf 28} (1972) 801]. 
The contribution of the lithium atoms to the electron-phonon coupling
is found to be negligible, while both boron and palladium atoms 
contribute equally strongly to the Hopfield parameter. 
There is a significant transfer of charge from lithium, almost the 
entire valence charge, to the B-Pd(Pt) complex. 
The electronic structure and superconducting properties of 
Li$_2$Pd$_3$B, thus, can be understood from the viewpoint of the 
compound being composed of a connected array of B-Pd tetrahedra 
decoupled from the backbone of Li atoms, which are connected by 
relatively short bonds.
Our results suggest that conventional $s$-wave electron-phonon 
interaction without explicit consideration of SO coupling can explain 
qualitatively the observed $T_c$ in Li$_2$Pd$_3$B. 
However, such an approach is likely to fail to describe 
superconductivity in Li$_2$Pt$_3$B. 
\end{abstract}

\end{frontmatter}

\section{Introduction}
The recently synthesized isotypic metal-rich ternary borides 
Li$_2$Pd$_3$B and Li$_2$Pt$_3$B \cite{eibenstein}, with
noncentrosymmetric crystal structures, have received 
considerable attention because of their superconducting properties 
\cite{togano,badica,sardar,mathi,nmr,x-ray}. 
The superconducting transition temperature $T_c$ of Li$_2$Pd$_3$B is 
about $8$ K, while its isotypic Pt-based counterpart Li$_2$Pt$_3$B 
shows a $T_c$ between $2$ and $3$ K \cite{badica2}. 
Badica and co-workers \cite{badica2} have been able to synthesize a
series of pseudo-binary solid solutions Li$_2$B(Pd$_{1-x}$Pt$_x$)$_3$ 
with $x$ varying from $0$ to $1$ and they report having observed 
superconductivity in the entire $x$-range. 
$T_c$ decreases monotonically from $\sim$ $8$ K as $x$ increases from 
$0$, dropping to $2.2$--$2.8$ K for $x = 1$. 
The electronic structure of the two end compounds of the solid 
solution Li$_2$Pd$_3$B and Li$_2$Pt$_3$B have been discussed by 
Chandra and co-workers \cite{mathi}. 
Sardar and Sa \cite{sardar} have used a three-dimensional single band 
$t$-$J$ model to discuss the superconductivity of Li$_2$Pd$_3$B, 
pointing out that the superconductivity in this compound arises from 
the Pd $4$\textit{d} electrons, and thus should be dominated by the 
strong correlation effects of narrow band Pd $4$\textit{d} electrons. 
The importance of the correlation effects of the $4$\textit{d} 
electrons for the superconducting properties of Li$_2$Pd$_3$B and 
Li$_2$Pt$_3$B was also mentioned by Chandra \textit{et al.} 
\cite{mathi}. 
On the other hand, Yokoya \textit{et al.} \cite{x-ray} report 
excellent agreement between their x-ray photoemission results for the 
valence band spectrum and standard band structure calculations, which 
can describe only weakly correlated materials. 
They conclude that the correlation effects play a negligible role for 
the physical properties of this superconductor, and that 
superconductivity in this material arises from the Pd $4$\textit{d} 
electrons hybridized with B $2$\textit{p} and Li $2$\textit{p} 
electrons. 
Absence of strong correlation effects in Li$_2$Pd$_3$B is also 
indicated in the \nuc{11}{B} NMR measurements in this compound by 
Nishiyama \textit{et al.} \cite{nmr}. 
Since the Pd(Pt) bands are almost full \cite{mathi}, the correlation 
effects are not expected to be strong.
The purpose of the present work is to examine the two isotypic 
superconductors Li$_2$Pd$_3$B and Li$_2$Pt$_3$B from the viewpoint of 
electron-phonon coupling and conventional (\textit{s}-wave) 
superconductivity, without explicit consideration of electron 
correlation effects and spin-orbit (SO) coupling. 
To this end, we use the rigid muffin-tin approach of Gaspari and 
Gyorffy \cite{gaspari} implemented within the linear muffin-tin orbital
(LMTO) method \cite{lmto} by Gl\"{o}tzel \textit{et al.} \cite{gloetz} 
and later by Skriver and Mertig \cite{sk-mt}.

The present paper is organized as follows. 
In section \ref{sec_electronicstructure} we discuss the electronic 
structure, charge transfer and bonding in the two compounds. 
In section \ref{sec_electron-phonon} we discuss the electron-phonon 
coupling. 
In particular we provide estimates of the Fermi-surface averaged 
electron-phonon matrix elements (Hopfield parameters), and 
electron-phonon coupling constants.  
Our results indicate that conventional electron-phonon coupling,
without consideration of SO effects, can describe the observed 
superconductivity in Li$_2$Pd$_3$B, but not in Li$_2$Pt$_3$B. 
There is some similarity in the superconducting properties of 
Li$_2$Pt$_3$B and another recently discovered noncentrosymmetric 
crystal CePt$_3$Si with strong SO coupling \cite{kirill}. 
This is discussed in section \ref{sec_electron-phonon}.
In section \ref{sec_conclusion} we present our conclusions.

\section{\label{sec_electronicstructure}Electronic Structure}

\begin{figure}
  \hfill
  \includegraphics[width=0.5\textwidth,clip=]{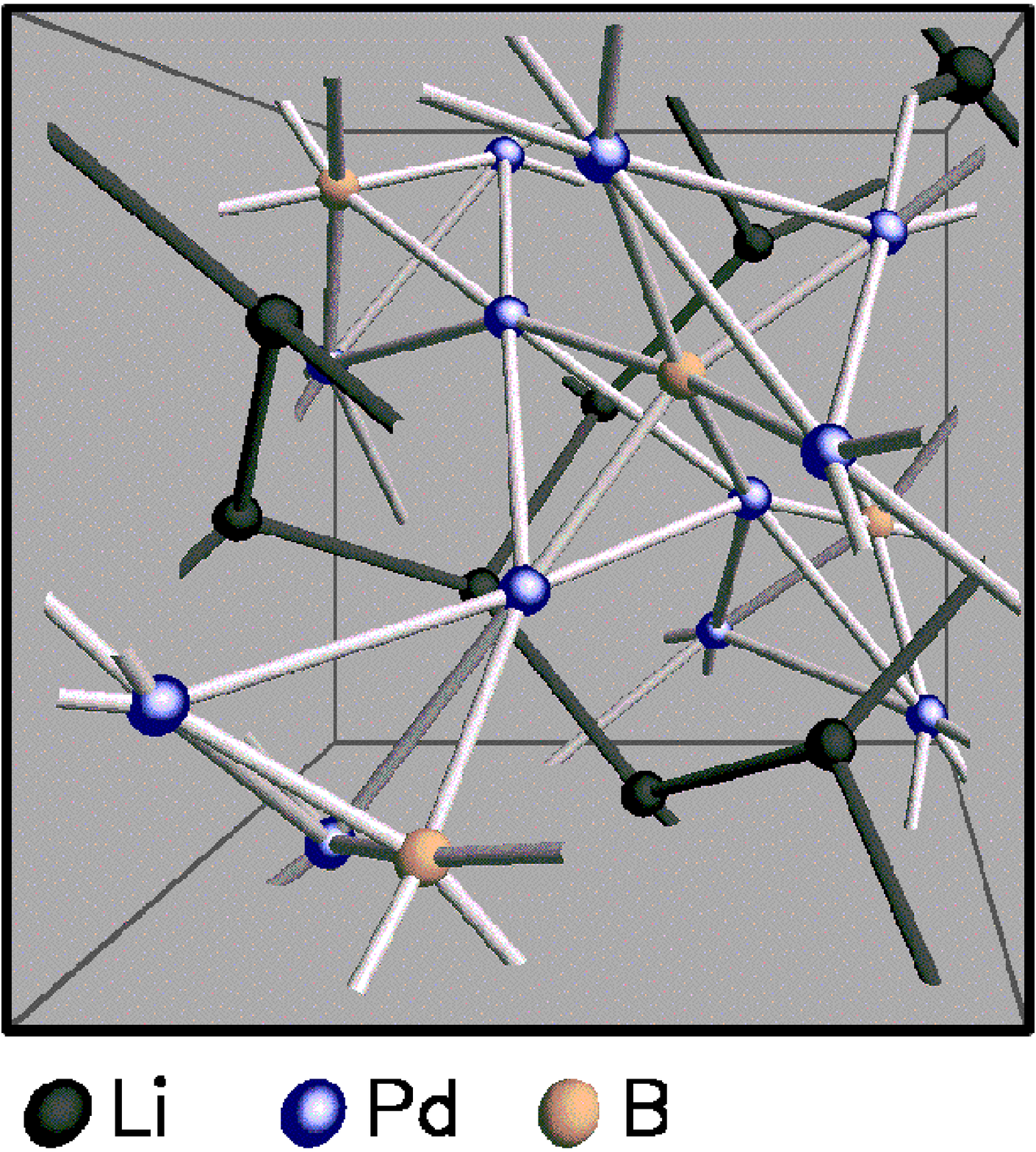}
  \hspace*{\fill}
  \caption{\label{fig_xyz}Unit cell of Li$_2$Pd$_3$B.}
\end{figure}

\begin{figure*}
  \includegraphics[width=0.49\textwidth,clip=]{fig2ab}\hfill
  \includegraphics[width=0.49\textwidth,clip=]{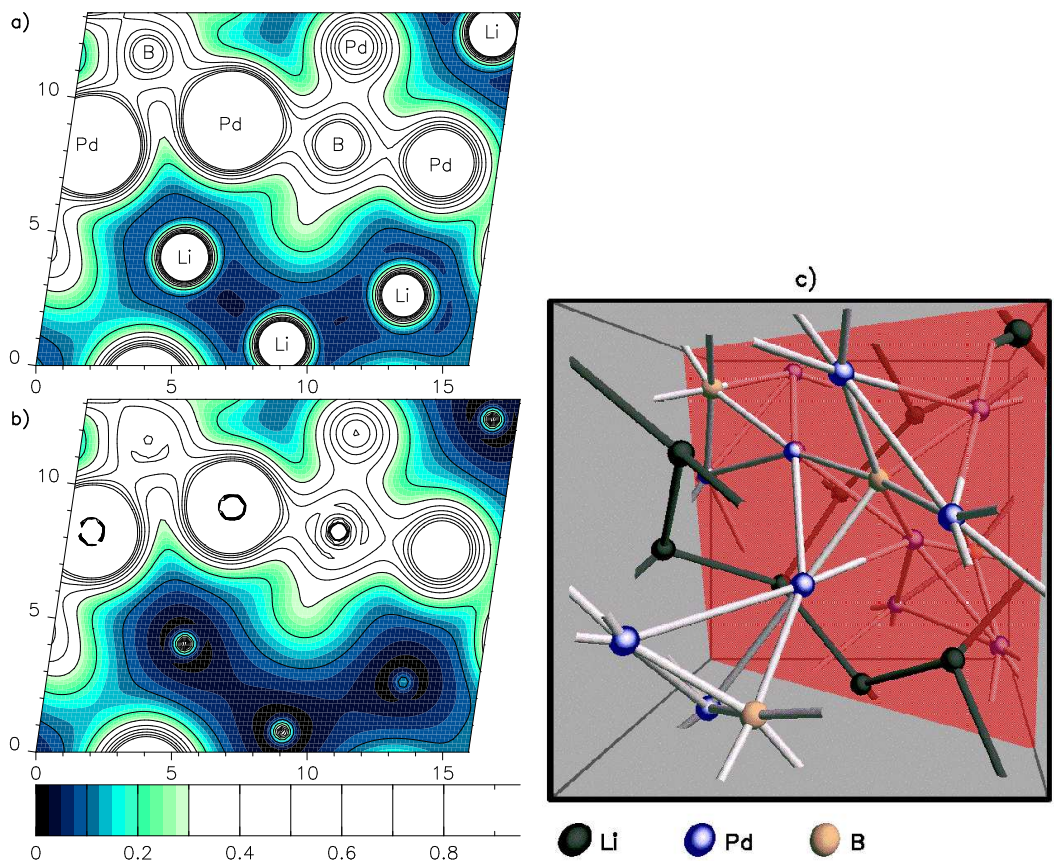}
  \caption{\label{fig_rho}(a) Total and (b) valence charge densities 
  of Li$_2$Pd$_3$B in (c) the plane that passes through
  the points ($0$,$0$,$0$), ($a/4$,$0$,$a$), and ($a$,$a$,$a$).}
\end{figure*}

\begin{figure}
  \hfill
  \includegraphics[width=0.5\textwidth,clip=]{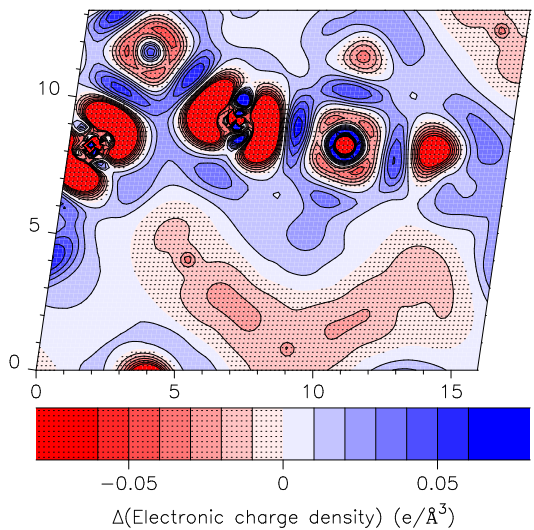}
  \hspace*{\fill}
  \caption{\label{fig_drho}Difference of the valence charge density of 
  Li$_2$Pd$_3$B and a superposition of atomic valence charge 
  densities.}
\end{figure}

The electronic structures of the two ternary borides Li$_2$Pd$_3$B and 
Li$_2$Pt$_3$B, which crystallize in the {\bf P}4$_3$32 structure 
(space group no.\ 212) with 4 formula units per unit cell 
\cite{eibenstein}, were calculated with the LMTO \cite{lmto} method 
using the atomic sphere approximation (ASA).
The lattice constants used were $12.762$ and $12.765$ $a_0$ for 
Li$_2$Pd$_3$B and Li$_2$Pt$_3$B, respectively.
The positions of the atoms were taken to be the same as given by 
Eibenstein and Jung \cite{eibenstein} and used in the electronic 
structure calculation of Chandra \textit{et al.} \cite{mathi}.
The atomic structure is shown in Fig.\ \ref{fig_xyz}.
The shortest bonds are between Pd and B ($4.0$ $a_0$) and Li and Li
($4.8$ $a_0$).
Each B atom is connected to six Pd atoms, and each Pd atom has two B
nearest neighbors. 
The structure of the Pd and B atoms can be described as a network of 
connected Pd$_3$B tetrahedra \cite{mathi}.
To show this, in Fig.\ \ref{fig_xyz} we have also 
drawn bonds for the shortest Pd-Pd distances ($5.3$ $a_0$), which
are however longer than some of the Pd-Li distances ($5.2$ $a_0$, not
shown).
The three-fold coordinated Li atoms form their own network 
(Fig.\ \ref{fig_xyz}).
Because the Li-Li distances are $16\%$ smaller than in bcc Li
\cite{sardar}, it has been anticipated \cite{sardar} that the Li atoms 
transfer their $2$\textit{s} electrons entirely to the Pd and B atoms.
The valence charge density of Li$_2$Pd$_3$B [Fig.\ \ref{fig_rho}(b), 
the total charge density is presented in Fig.\ \ref{fig_rho}(a)],
shows indeed that there is very little electronic valence charge 
present near the Li atoms.
To see the effect of bonding it is often useful to subtract the atomic
valence charge densities from the total valence charge density.
The result is shown in Fig.\ \ref{fig_drho}.
Again it is clear that the network of Li atoms is depleted of charge.
In addition, a strong increase in the charge density is found between
the B and Pd atoms, which indicates covalent bonding.
The increase in charge density between the Pd atoms is less pronounced,
suggesting that B-Pd bonds are stronger than Pd-Pd bonds.

\begin{figure}
  \hfill
  \includegraphics[width=0.8\textwidth]{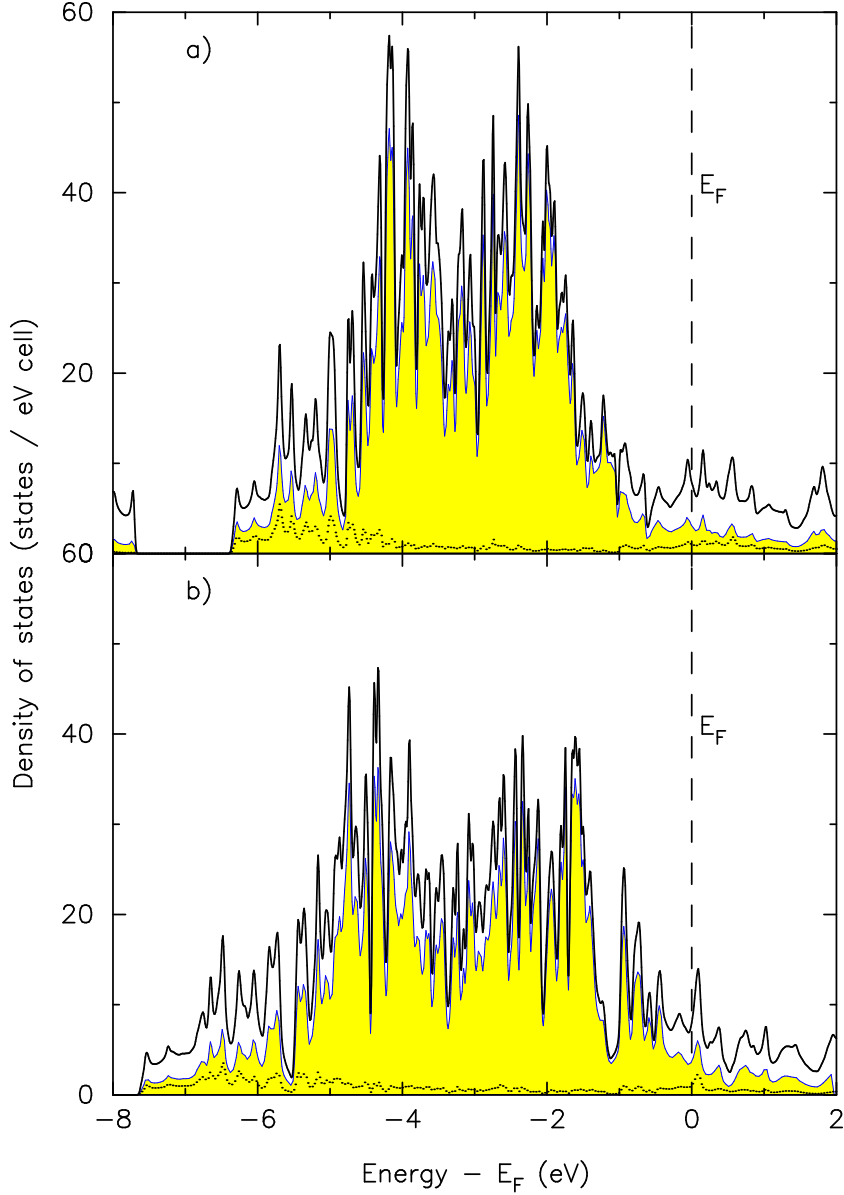}
  \hspace*{\fill}
  \caption{\label{fig_dos}Densities of states of (a) Li$_2$Pd$_3$B
  and (b) Li$_2$Pt$_3$B (thick solid curves). 
  Partial Pd and Pt-\textit{d} (thin curves, shadings) and 
  B-\textit{p} (dotted curves) densities of states are also shown.}
\end{figure}

We have carried out the LMTO calculation without empty spheres (which 
a more rigorous calculation would have required) using the $spdf$ 
basis and with the $f$-orbitals downfolded. 
Even though our results are not the best possible results within the 
LMTO-ASA scheme, a comparison with more rigorous full-potential linear 
augmented plane waves plus local orbitals (FP-LAPW+lo) results 
presented by Chandra \textit{et al.} \cite{mathi} shows that our 
results are of acceptable accuracy. 
The densities of states shown in Fig.\ \ref{fig_dos} 
agree very well with  the  results of Chandra \textit{et al.} 
\cite{mathi}.
The agreement of our calculated densities of states with another set of
FP-LAPW results of Yokoya {\it et al.} \cite{x-ray} is also very good. 
In Fig.\ \ref{fig_dos} among the various partial densities of states,
only the Pd(Pt)-$d$ and B-$p$ densities of states are shown, since 
these have the largest values at the Fermi level and as a result 
electron-phonon interaction is dominated by Pd(Pt) $d$-$f$ and B 
$p$-$d$ scattering (section \ref{sec_electron-phonon}).
Note that the density of states per atom is almost the same for
Pd-$d$ (3.34 States/(Ry atom)) and B-$p$ states (3.23 States/(Ry atom))
in Li$_2$Pd$_3$B. 
In Li$_2$Pt$_3$B these numbers corresponding to Pt-$d$ and B-$p$ 
states are 4.2 States/(Ry atom) and 3.72 States/(Ry atom), 
respectively.

The main features of our LMTO calculations that are used in estimating 
the electron-phonon matrix elements in section 
\ref{sec_electron-phonon} are the atom and orbital resolved densities 
of states, listed in Table \ref{table1}.   
\begin{table}
  \caption{Partial $l$ and atom resolved densities of states, 
  including both spins, at the Fermi energy in states / Ry for 
  $24$ atom unit cells of Li$_2$Pd$_3$B and Li$_2$Pt$_3$B. 
  The numbers shown are for $12$ Pd(Pt), $8$ Li and $4$ B atoms. 
  The total Fermi level DOS, $n(E_F)$, is $102.75$ states / Ry / cell 
  for Li$_2$Pd$_3$B and $107.8$ states/Ry/cell for Li$_2$Pt$_3$B.}
  \label{table1}
  \begin{tabular}{cccccc}
    \hline
    compound      & atom & $n_s$ &  $n_p$  &  $n_d$  &  $n_f$  \\
    \hline
    Li$_2$Pd$_3$B & Pd & $7.132$ & $18.54$ & $40.12$ & $1.162$ \\
                  & Li & $6.000$ & $11.53$ & $2.566$ & $0.481$ \\
                  & B  & $0.887$ & $12.92$ & $1.084$ & $0.333$ \\
    Li$_2$Pt$_3$B & Pt & $7.499$ & $17.38$ & $50.47$ & $1.980$ \\
                  & Li & $3.284$ & $6.791$ & $2.586$ & $0.709$ \\
                  & B  & $0.837$ & $14.89$ & $1.093$ & $0.361$ \\
    \hline
  \end{tabular}
\end{table}
In addition, the potentials at the sphere boundaries and LMTO 
potential parameters calculated with the LMTO reference energies set 
at the Fermi energy also enter the expression for the electron-phonon 
matrix elements, as discussed in section \ref{sec_electron-phonon}.

\section{\label{sec_electron-phonon}Electron-phonon coupling and 
superconductivity}

According to McMillan's strong coupling theory \cite{mcmillan} the 
electron-phonon coupling constant $\lambda_{e-ph}$ for a one-component
system, i.e elemental solid, can be written as
\begin{equation}
  \lambda_{e-ph} = \frac{n(E_F) \langle I^2 \rangle}{M \langle 
  \omega^2 \rangle} \;,
\end{equation}
where $M$ is the atomic mass, $\langle \omega^2 \rangle$ is the 
renormalized phonon frequency, squared and averaged according to the 
prescriptions in Ref. \cite{mcmillan}, $n(E_F)$ is the density of 
states for one type of spin at the Fermi energy $E_F$, and $\langle 
I^2 \rangle$ is the square of the electron-phonon matrix element 
averaged over the Fermi surface. 
Gaspari and Gyorffy \cite{gaspari} constructed a theory to calculate 
the quantity  $\langle I^2 \rangle$ on the assumption that the 
additional scattering of an electron caused by the displacement of an 
atom (ion) is dominated by the change in the local potential. 
Within the rigid muffin-tin approximation used by Gaspari and Gyorffy 
\cite{gaspari} the spherically averaged part of the Hopfield parameter
$\eta = n(E_F) \langle I^2 \rangle$ 
can be written as (in atomic Rydberg units)
\begin{equation}
  \eta = 2 n(E_F) \sum_l (l + 1) M^2_{l,l+1} \frac{f_l}{2 l + 1}
  \frac{f_{l + 1}}{2 l + 3}\;,
\end{equation}
where $f_l$ is a relative partial state density,
\begin{equation}
  f_l = \frac{n_l \left( E_F \right)}{n(E_F)}
\end{equation}
and $M_{l,l+1}$ is the electron-phonon matrix element obtained from 
the gradient of the potential and the radial solutions $R_l$ and 
$R_{l+1}$ of the Schr{\"{o}}dinger equation evaluated at the Fermi 
energy. 
Gaspari and Gyorffy \cite{gaspari} derived an expression for 
$M_{l,l+1}$ using the rigid muffin-tin approximation in terms of 
partial wave phase shifts.
Gl\"{o}tzel {\it et al.} \cite{gloetz} and Skriver and Mertig 
\cite{sk-mt}, using the LMTO \cite{lmto} method, expressed this 
quantity in terms of the logarithm derivative $D_l(E_F)$ of the radial
solution at the sphere boundary, with the reference energy $E_{\nu}$ 
set at the Fermi energy $E_F$:
\begin{align}
  M_{l,l+1} = - \phi_l(E_F) \phi_{l+1}(E_F) \{ & \left[ D_l(E_F) - l 
  \right] \left[ D_{l+1}(E_F) + l + 2 \right] \nonumber\\
  & + \left[ E_F - V(S) \right] S^2 \},
\end{align}
where $S$ is the sphere radius, $V(S)$ is the one-electron potential 
and $\phi_l(E_F)$ the sphere-boundary amplitude of the $l$ partial 
wave evaluated at the Fermi level.

The extension of the Gaspari and Gyorffy scheme \cite{gaspari} to 
alloys has been discussed by several authors 
\cite{gomersall-gyorffy,klein-papa,papa-klein,klei-etal,jarlborg,mazin,rainer}.
To obtain an estimate of the Hopfield parameter and gain some insight
into electron-phonon coupling in the two isotypic compounds of 
interest, in this paper we adopt the following simplified approach.
Since $\langle I^2 \rangle$, in the rigid muffin-tin or rigid atomic 
sphere approximation, is an atomic quantity, it can be calculated for 
each atom in the alloy using the formula given by Skriver and Mertig 
\cite{sk-mt}, by considering atom-resolved partial densities of 
states. 
For atom A
\begin{equation}
  \langle I_A^2 \rangle = 2 \sum_l (l + 1) M^2_{\left( A, \left(l, 
  l + 1 \right) \right)} \frac{f^A_l}{2 l + 1} \frac{f^A_{l + 1}}
  {2 l + 3},
\end{equation}
where $ f^A_l = \frac{n^A_l \left( E_F \right)}{n(E_F)} \;.$
If there are $N_A$ atoms of type A in the unit cell, then 
$n(E_F) = \sum_{A,l} N_A n^A_l \left( E_F \right)$ and the Hopfield 
parameter $\eta$ can be estimated  from
\begin{equation}
  \eta = \sum_A \eta_A = n(E_F) \sum_A N_A \langle I_A^2 \rangle\;,
\end{equation}
an expression, which is independent of the size of the unit cell 
chosen.
This expression ignores possible
contributions from cross-terms involving more than one atom.
The quantities $M^2_{\left( A, \left( l, l + 1 \right) \right)}$ can 
be calculated from the sphere-boundary potentials and the LMTO-ASA 
potential parameters evaluated with the reference energies set equal 
to the Fermi energy.

The values for the total and partial densities of states for 
Li$_2$Pd$_3$B and Li$_2$Pt$_3$B are shown in Table \ref{table1}.  
In Table \ref{table2} 
\begin{table}
  \caption{Bulk modulus $B$ in GPa, Debye temperature $\Theta_D$ in K, 
  atom-resolved and total Hopfield parameters given by Eqs.\ (5) and 
  (6) in eV / {\AA}$^2$ for Li$_2$Pd$_3$B and Li$_2$Pt$_3$B.
  $\eta$ denotes the total Hopfield parameter.}
  \label{table2}
  \begin{tabular}{ccccccc}
    \hline
    compound & $B$ & $\Theta_D$ & $\eta_{\mathrm{Pd}}$  
             & $\eta_{\mathrm{B}}$ & $\eta_{\mathrm{Li}}$ & $\eta$ \\
    \hline
    Li$_2$Pd$_3$B & $166$ & $371$ & $1.30$ & $1.10$ & $0.19$ & $2.59$\\
    Li$_2$Pt$_3$B & $196$ & $303$ & $2.67$ & $3.69$ & $0.08$ & $6.44$\\
    \hline
  \end{tabular}
\end{table}
the atom-resolved and the total Hopfield parameters, $\eta_A$ and 
$\eta$, obtained by using Eqs.\ (5) and (6) are listed. 
We also list the calculated bulk moduli of the two compounds, which 
can be used to obtain estimates of the Debye temperatures and average 
phonon frequencies for these solids.
The analysis of the electron-phonon coupling and critical temperatures
in the two compounds is based on the quantities listed in Tables 
\ref{table1} and \ref{table2}.  

\subsection{Li$_2$Pd$_3$B}
 
For Li$_2$Pd$_3$B the Hopfield parameter calculated according to 
Eqs.\ (5) and (6) is $2.59$ eV / {\AA}$^2$.
Note that the boron contribution is only $15\%$ less than the 
palladium contribution.
Most dominant contributions come from Pd $d$-$f$ ($0.98$ eV / 
{\AA}$^2$) and B $p$-$d$ ($0.65$ eV / {\AA}$^2$) scattering.  
Although double in number than the B atoms, all Li atoms together 
contribute only $7\%$ to the Hopfield parameter $\eta$. 
$95\%$ of the contribution of the Li atoms is from the $s-p$ channel.  

A few comments regarding the decomposition of the Hopfield parameter 
into atom-resolved parts and various partial wave channels in an alloy
are in order. 
Because certain potential parameters depend on the ratio $s/W$ (or 
$\left( s / W \right)^{2 l + 1}$), where $s$ is the sphere radius
for the atom and $W$ is the average Wigner-Seitz radius of the alloy,
the atom-resolved parameters may differ for different choices of the 
sphere radii, all chosen within the allowable sphere overlaps. 
The total Hopfield parameter is expected to be independent of the
choice of the sphere radii, if these are chosen to keep the ASA errors
as small as possible, but the partial-wave resolved parameters 
(i.e, $s$-$p$, $p$-$d$ etc.) are dependent on the ratio $s/W$.
Ideally, the LMTO calculation for the two solids Li$_2$Pd$_3$B and 
Li$_2$Pt$_3$B should have been carried out with empty spheres in order
to reduce ASA errors. 
This would have given an improved value of the total Hopfield 
parameter, including contributions from both atomic and empty spheres.
Despite some ASA errors present in our results, we expect the trends 
revealed by them to be correct.

In the simplest approximation, assuming isotropic electron-phonon 
coupling, one can attempt to estimate $\lambda_{e-ph}$ from $\eta$ 
using Eq.\ (1), assuming that $M$ is the concentration average of the 
mass of the component atoms in the alloy and $\omega$ is the average 
phonon frequency for the alloy. 
To estimate the average phonon frequency we use the prescription of 
Moruzzi, Janak and Schwarz \cite{MJS}, relating the Debye temperature 
$\Theta_D$ to the bulk modulus, atomic mass and the average 
Wigner-Seitz radius:
\begin{equation}
  \Theta_D = 41.63 \sqrt{\frac{r_0 B}{M}}\;,
\end{equation}
where $r_0$ is the average Wigner-Seitz radius in atomic units, $B$ is
the bulk modulus in kbar and $M$ is the atomic mass. 
Although Moruzzi {\it et al.} \cite{MJS} verified the validity of 
this expression for elemental metallic solids, we assume its validity 
for the alloy by considering $M$ to be the concentration average of 
the masses of the component atoms. 
From the LMTO calculation of total energy versus volume we find the 
bulk modulus $B$ of Li$_2$Pd$_3$B to be $166$ GPa, a value consistent 
with the bulk moduli of elemental solids Pd ($B = 180.8$ GPa), B 
($B = 178$ GPa) and Li ($B = 11.6$ GPa) \cite{kittel}.
Using $166$ GPa for $B$ in the above equation we find $\Theta_D = 371$
K. 
Using the empirical estimate $\sqrt{ \langle \omega^2 \rangle} = 0.69
\Theta_D$ \cite{sk-mt} yields an average phonon frequency of $5.3$ THz.
The use of Eq.\ (1) then yields $\lambda_{e-ph} = 0.39 K \sim 0.4 K$, 
a value not large enough to explain  the observed superconductivity of 
Li$_2$Pd$_3$B at $\sim 8$ K. 
This could have been foreseen from the small value, $2.59$ eV / 
{\AA}$^2$, of $\eta$ itself. 
The value of $\eta$ for niobium, which has a comparable $T_c$ of $9$ K,
is in the range $5.4$--$7.6$ eV / {\AA}$^2$ \cite{papa-etal}. 
However, in view of the uncertainties, not only in the frequency 
value, but the validity of Eq.\ (1) itself, this value of 
$\lambda_{e-ph}$ cannot be expected to be more than a coarse 
approximation.
It is known that even for one-component systems the rigid muffin-tin 
approximation usually gives values  of the Hopfield parameter that are 
lower than those obtained via more rigorous calculations of the 
electron-phonon interaction (see Table II. in Ref. \cite{bose-etal}). 
For the solids under consideration with a large number of optical 
phonon branches, the extent of underestimation is probably more 
severe.  
What is learnt from the above rigid muffin-tin (or atomic sphere) 
calculation is that the contribution to the electron phonon coupling 
from both boron and palladium should be equally important, and that 
from the lithium atoms should be negligible.

Following the prescription of Gomersall and Gyorffy 
\cite{gomersall-gyorffy,klein-papa} one could try to analyse the 
electron-phonon coupling by writing
\begin{equation}
  \lambda_{e-ph} = \lambda_{Pd} + \lambda_B = \frac{\eta_{Pd}}{M_{Pd}
  \langle \omega^2 \rangle_{Pd}} + \frac{\eta_B}{M_B \langle \omega^2
  \rangle_B}\;,
\end{equation}
where the subscripts Pd and B refer to atom-resolved quantities. 
This decomposition is based on the large difference between the Pd and
B masses, and assumes that Li atoms do not contribute significantly to
the electron-phonon coupling, as supported by the LMTO results. 
The heavy palladium atom vibration is assumed to couple to the 
electrons through zone boundary acoustic phonons and the lighter boron
atom vibration is assumed to couple to electrons via the optical modes.
It is reasonable to assume that in the acoustic modes only the 
palladium  atoms vibrate, and in the optical modes only the lighter 
boron atoms vibrate.
Instead of guessing the average phonon frequencies in the above 
equation, we can estimate the ratio $\lambda_{Pd} / \lambda_B$ by 
assuming that the average optical phonon frequency is, to take an 
example, $2.5$ times higher than the average zone boundary acoustic 
phonon frequency. 
With the values of $\eta_{Pd}$ and $\eta_B$ quoted in Table 
\ref{table2}, we then find $\lambda_{Pd} = 0.75 \lambda_B$, i.e. if 
the above scenario holds, then the boron contribution to the 
electron-phonon coupling is larger than the Pd contribution. 
For $\lambda_{Pd}$ to be equal to or greater than  $\lambda_B$, in 
this scenario, the average optical phonon frequency needs to be about 
$2.9$ times or higher than the average zone boundary acoustic phonon 
frequency. 
From the consideration of the linear specific heat coefficient 
\cite{togano-takeya}, Lee and Pickett \cite{lee-pickett} have obtained
$\lambda_{e-ph} = 0.74$. 
With $\lambda_{Pd} = 0.75 \lambda_B$, this would yield $\lambda_{Pd} 
= 0.32$ and $\lambda_B = 0.42$. 
Note that the value $\lambda_{Pd} = 0.32$ is not unreasonable, as a 
full-potential LMTO linear response calculation by Savrasov and 
Savrasov \cite{savrasov} yields $\lambda_{Pd} = 0.35$ in elemental fcc
Pd. 
Incidentally, the use of the McMillan expression \cite{allen-mitro}:
\begin{equation}
  T_{c} = \frac{\Theta_D}{1.45} \exp \left \{ - \frac{1.04 \left( 1 +
  \lambda_{e-ph} \right)}{ \lambda_{e-ph} - \mu^{\ast}(1 + 0.62 
  \lambda_{e-ph})} \right\}, \label{mcm}
\end{equation}
with the Coulomb pseudopotential $\mu^{\ast}$ chosen to be $0.1$, as 
it is often done, $\Theta_D= 371$ K (Table \ref{table2}), and 
$\lambda_{e-ph} =0.74$ yields $T_c \sim 12 $ K. 

The above discussion points out that a description based on 
conventional ($s$-wave, spin singlet Cooper pairs) electron-phonon 
interaction should be applicable to Li$_2$Pd$_3$B. 
Of course a more rigorous calculation of the electron-phonon 
interaction is needed. 
Such a calculation is likely to reveal negligible coupling of lithium 
atom vibrations to the electron states at the Fermi level, while the 
vibrations of the palladium and boron atoms should couple strongly.

According to Yuan \textit{et al.} \cite{yuan} temperature dependence 
of the penetration depth in Li$_2$Pd$_3$B could be fit by a two-gap 
BCS model with a small ($3.2$ K) and a large ($11.5$ K) gap. 
Thus the claim is that Li$_2$Pd$_3$B is  a two-gap BCS superconductor 
just like MgB$_2$ \cite{soma,mazin2,boza1}. 
The Fermi surface plots presented by Chandra {\it et al.} \cite{mathi}
show a Fermi surface with several sheets: a central large sheet and 
several smaller ones originating from the four bands crossing the 
Fermi level. 
A two or multi band superconductivity is thus conceivable in this 
material just as in MgB$_2$ with two distinct sheets in the Fermi 
surface \cite{jens,kong}.
However, $T_c$ may still be dominated by a single band \cite{boza2}, 
particularly in the absence of strong interband scattering. 

\subsection{Li$_2$Pt$_3$B and the importance of SO coupling}

The inadequacy of the above approach to describe the electron-phonon 
coupling in Li$_2$Pt$_3$B becomes clear from the much larger value of 
the Hopfield parameter obtained for this compound and listed in Table 
\ref{table2}. 
Note that the unusually large contribution from boron is partly due to 
larger difference between the boron sphere radius $s$ and the average 
Wigner-Seitz radius $W$. 
A calculation including empty spheres to reduce ASA errors would have
changed the numbers in Table \ref{table2} somewhat, without changing
the results qualitatively.
Even with the reduced value of $\Theta_D = 303 K$ (Table \ref{table2}),
the much larger value of $\eta$ would yield a value for 
$\lambda_{e-ph}$ larger than that for Li$_2$Pd$_3$B, implying a $T_c$ 
that is actually higher than in Li$_2$Pt$_3$B.
This is in clear contradiction to the experimental observation that 
$T_c$ decreases monotonically as the Pt concentration increases from 
zero to unity

The most serious drawback of the above analysis is the neglect of 
SO coupling, which is more pronounced in Li$_2$Pt$_3$B than in 
Li$_2$Pd$_3$B.  
SO coupling, in the absence of the center of inversion symmetry in 
these two compounds, can result in substantial splittings of the spin 
degenerate bands considered in this work. 
In a  recent submission, Lee and Pickett \cite{lee-pickett} discuss SO
coupled bands in both compounds: splitting of the spin degenerate 
bands is found to be as large as $30$ meV and $200$ meV at the Fermi 
level in Li$_2$Pd$_3$B and Li$_2$Pt$_3$B, respectively. 
The SO splitting of $30$ meV or $348$ K in Li$_2$Pd$_3$B is less than 
the estimated Debye temperature $371$ K, the cut-off energy of the
electron-phonon interaction. 
In Li$_2$Pt$_3$B the SO splitting is $200$ meV or $2321$ K, an order 
of magnitude larger than the estimated Debye temperature $303$ K. 
It is clear that SO coupling would strongly alter the symmetry of the 
Cooper pairs, and the nature of the superconducting state in 
Li$_2$Pt$_3$B \cite{kirill,gorkov,frigeri,sergienko}.

Yuan {\it et al.} \cite{yuan} have claimed evidence of line of nodes 
in the energy gap in Li$_2$Pt$_3$B. 
Such a line of zeros in the energy gap is also seen in CePt$_3$Si 
\cite{izawa}, another noncentrosymmetric crystal with strong SO 
splitting of the spin degenerate bands \cite{kirill}. 
The existence of the line of nodes is not a symmetry-imposed 
requirement and remains to be explained for both Li$_2$Pt$_3$B and 
CePt$_3$Si.  
CePt$_3$Si is a heavy fermion superconductor ($T_c = 0.75$ K), so 
the mechanism of the pair formation is unclear. 
Li$_2$Pt$_3$B is most probably an electron-phonon superconductor, 
where the electron-phonon coupling needs to be determined from the SO 
coupled electron wavefunctions and is expected to differ strongly from
that determined from the spin degenerate wavefunctions.

\section{\label{sec_conclusion}Conclusions}

The results presented in this work are not rigorous, and should only 
be used as a guide to understand the relative strengths of coupling of 
the vibration of the different atoms to the electron states at the 
Fermi level.
Based on these results, a description of superconductivity assuming 
conventional $s$-wave electron-phonon coupling and via a single 
electron-phonon coupling parameter appears possible in Li$_2$Pd$_3$B. 
The inclusion of SO coupling in the calculation of electron-phonon 
interaction should lead to improvement in results.         
For Li$_2$Pt$_3$B the SO effects are an order of magnitude stronger 
and do need to be included in the calculation of the electron-phonon
interaction in order to explain the lower $T_c$ in this compound. 

The trends and the conclusions reached for Li$_2$Pd$_3$B in this work 
should be tested via rigorous calculation of the phonon frequencies 
and electron-phonon interaction. 
This is particularly important in view of the large number of phonon
branches in this material. 
The results of this paper indicate that only the modes where the boron
and the palladium atoms vibrate should couple to the electrons. 
The electron-phonon coupling should be equally strong for the modes
where essentially the Pd atoms or the B atoms vibrate.
According to our analysis the modes involving the vibrations of the Li
atoms should show negligible coupling to electron states at the Fermi 
level.
This is also suggested by our finding that almost the entire valence 
charge of lithium is transferred to the Pd-B complex (section 
\ref{sec_electronicstructure}). 
Recent work of Lee and Pickett \cite{lee-pickett} also supports this 
conclusion.
First principles linear response calculations of the phonons and 
electron-phonon coupling to verify the results of the present work are
currently in progress.

Due to the strong SO coupling in Li$_2$Pt$_3$B the properties of the 
superconducting state is expected to deviate strongly from that 
described by spin-degenerate wavefunctions 
\cite{kirill,gorkov,frigeri,sergienko}. 
However, superconductivity should still be of the electron-phonon type
(possibly also spin-singlet type, as suggested by Lee and Pickett 
\cite{lee-pickett}). 
The experimental result that $T_c$ decreases monotonically in solid 
solutions Li$_2$B(Pd$_{1-x}$Pt$_x$)$_3$ as the Pt concentration 
increases from zero to $1$ indicates that the mechanism of electron
pair formation remains unchanged, i.e. of the electron-phonon type.

\begin{ack}
Financial support for this work was provided by Natural Sciences and 
Engineering Research Council of Canada. 
SKB gratefully acknowledges helpful discussion with O.\ Jepsen, 
B.\ Mitrovi\'{c} and K.\ V.\ Samokhin.
\end{ack}


\begin{thebibliography}{00}




\bibitem{eibenstein} U.\ Eibenstein and W.\ Jung, 
                     J.\ Solid State Chem.\ \textbf{133} (1997) 21.
\bibitem{togano} K.\ Togano, P.\ Badica, Y.\ Nakamori, S.\ Orimo, 
                 H.\ Takeya and K.\ Hirata, 
                 Phys.\ Rev.\ Lett.\ \textbf{93} (2004) 247004.
\bibitem{badica} P.\ Badica, T.\ Kondo, T.\ Kudo, Y.\ Nakamori, 
                 S.\ Orimo and K.\ Togano, cond-mat/0404257.
\bibitem{mathi} S.\ Chandra, S.\ Mathi Jaya and V.C.\ Valsakumar, 
                cond-mat/0502525.
\bibitem{sardar} M.\ Sardar and D.\ Sa, cond-mat/040168.
\bibitem{nmr} M.\ Nishiyama, Y.\ Inada and G.-q.\ Zheng, 
              Phys.\ Rev.\ B {\bf 71} (2005) 220505(R); 
              cond-mat/0506191 v1.
\bibitem{x-ray} T.\ Yokoya, T.\ Muro, I.\ Hase, H.\ Takeya, 
                K.\ Hirata and K.\ Togano, 
                Phys.\ Rev.\ B {\bf 71} (2005) 092507.
\bibitem{badica2} P.\ Badica, T.\ Kondo and K.\ Togano, 
                  J.\ Phys.\ Soc.\ Jpn {\bf 74} (2005) 1014.
\bibitem{gaspari} G.D.\ Gaspari and B.L.\ Gyorffy, 
                  Phys.\ Rev.\ Lett.\ {\bf 28} (1972) 801.
\bibitem{lmto} O.K.\ Andersen, Phys.\ Rev.\ B {\bf 8} (1975) 3060; 
               O.K.\ Andersen, O.\ Jepsen, M.\ Sob in 
               \textit{Electronic structure and its applications}, 
               edited by M.\ Yossouff, Lecture Notes in Physics, 
               v.\ 283 (Springer, Berlin, 1987) p.\ 1--57; 
               O.K.\ Andersen, O.\ Jepsen, and D.\ Gl\"{o}tzel, in
               \textit{Highlights of Condensed Matter Theory}, 
               edited by F.\ Bassani \textit{et al.} 
               (North-Holland, Amsterdam, 1985), p.\ 59; 
               H.L.\ Skriver, \textit{The LMTO Method} 
               (Springer, Berlin, 1984).
\bibitem{gloetz} D.\ Gl\"{o}tzel, D.\ Rainer and H.R.\ Schober, 
                 Z.\ Phys.\ B {\bf 35} (1979) 317.
\bibitem{sk-mt} H.L.\ Skriver and I.\ Mertig, 
                Phys.\ Rev.\ B {\bf 32} (1985) 4431.
\bibitem{kirill} K.V.\ Samokhin, E.S.\ Zijlstra and S.K.\ Bose, 
                 Phys. Rev. B {\bf 69} (2004) 094514; 
                 {\bf 70} (2004) 069902(E).
\bibitem{mcmillan} W.L.\ McMillan, Phys.\ Rev.\ {\bf 167} (1968) 331.
\bibitem{gomersall-gyorffy} I.R.\ Gomersall and B.L.\ Gyorffy, 
                            J.\ Phys.\ F {\bf 3} (1973) L138.
\bibitem{klein-papa} B.M.\ Klein and D.A.\ Papaconstantopoulos, 
                     Phys.\ Rev.\ Lett.\ {\bf 32} (1974) 1193.
\bibitem{papa-klein} D.A.\ Papaconstantopoulos and B.M.\ Klein, 
                     Phys. Rev. Lett. {\bf 35} (1975) 110.
\bibitem{klei-etal} B.M.\ Klein, W.E.\ Pickett, 
                    D.A.\ Papaconstantopoulos and L.L.\ Boyer, 
                    Phys.\ Rev.\ B {\bf 27} (1983) 6721.
\bibitem{jarlborg} T.\ Jarlborg, A.\ Junod and M.\ Peter, 
                   Phys.\ Rev.\ B {\bf 27} (1983) 1558.
\bibitem{mazin} I.I.\ Mazin and S.N.\ Rashkeev, 
                Phys.\ Rev.\ B {\bf 42} (1990) 366.
\bibitem{rainer} D.\ Rainer, 
                 Prog.\ Low Temp.\ Phys., vol.\ X (1986) 371.
\bibitem{papa-etal} D.A.\ Papaconstantopoulos, L.L.\ Boyer, 
                    B.M.\ Klein, A.R.\ Williams, V.L.\ Morruzzi
                    and J.F.\ Janak, 
                    Phys.\ Rev.\ B {\bf 15} (1977) 4221.
\bibitem{MJS} V.L.\ Moruzzi, J.F.\ Janak and K.\ Schwarz, 
              Phys.\ Rev.\ B {\bf 37} (1988) 790.
\bibitem{kittel} C. Kittel, {\it Introduction to Solid State Physics}, 
                 7$^{th}$ edn.\ (John Wiley, 1996), p.\ 59.
\bibitem{bose-etal} S.K.\ Bose, O.V.\ Dolgov, J.\ Kortus, O.\ Jepsen
                    and O.K.\ Andersen, 
                    Phys.\ Rev.\ B {\bf 67} (2003) 214518.
\bibitem{togano-takeya} K.\ Togano and H.\ Takeya, unpublished results
                        communicated to W.E.\ Pickett 
                        (Cond-mat/0507105 v1).
\bibitem{lee-pickett} K.-W.\ Lee and W.E.\ Pickett, 
                      cond-mat/0507105 v1.
\bibitem{savrasov} S.Y.\ Savrasov and D.Y.\ Savrasov, 
                   Phys.\ Rev.\ B {\bf 54} (1996) 16487.
\bibitem{allen-mitro} P.B.\ Allen and B.\ Mitrovi\'{c}, 
                      {\it Solid State Physics}, edited by
                      H.\ Ehrenreich, F.\ Seitz and D.\ Turnbull 
                      (Academic, New York 1982), vol.\ 37, p.\ 1.
\bibitem{yuan} H.Q.\ Yuan, D.\ Vandervelde, M.B.\ Salamon, P.\ Badica
               and K.\ Togano, cond-mat/0506771.
\bibitem{soma} S.\ Soma {\it et al.}, 
               Nature (London) {\bf 423} (2003) 65.
\bibitem{mazin2} I.I.\ Mazin and V.P.\ Antropov, 
                 Physica C {\bf 385} (2003) 49.
\bibitem{boza1} B.\ Mitrovi\'{c}, 
                J.\ Phys.: Condens.\ Matter {\bf 16} (2004) 9013.
\bibitem{jens} J.\ Kortus, I.I.\ Mazin, K.D.\ Belashchenko, 
               V.P.\ Antropov and L.L.\ Boyer, 
               Phys.\ Rev.\ Lett.\ \textbf{86} (2001) 4656.
\bibitem{kong} Y.\ Kong, O.V.\ Dolgov, O.\ Jepsen and O.K.\ Andersen, 
               Phys.\ Rev.\ B \textbf{64} (2001) 020501(R).
\bibitem{boza2} B.\ Mitrovi\'{c}, 
                Eur.\ Phys.\ J.\ B {\bf 38} (2004) 451.
\bibitem{gorkov} L.P.\ Gorkov and E.I.\ Rashba, 
                 Phys. Rev. Lett. {\bf 87} (2001) 037004.
\bibitem{frigeri} P.A.\ Frigeri {\it et al.}, 
                  Phys. Rev. Lett. {\bf 92} (2004) 097001.
\bibitem{sergienko} I.A.\ Sergienko and S.H.\ Curnoe, 
                    Phys.\ Rev.\ B {\bf 70} (2004) 214510.
\bibitem{izawa} K.\ Izawa {\it et al.}, 
                Phys.\ Rev.\ Lett.\ {\bf 94} (2005) 197002.
\end{thebibliography}
\end{document}